\documentclass[showpacs,aps,superscriptaddress,prb,twocolumn,amsmath]
{revtex4}
\usepackage{txfonts}
\usepackage{color}
\usepackage{amssymb}
\usepackage{graphicx}
\usepackage{dcolumn}
\usepackage{bm}
\usepackage{hyperref}
\usepackage{color}

\begin{document}
\preprint{APS/123-QED}
\title{Direct and quasi-direct band gap silicon allotropes with remarkable stability}
\author{Chaoyu He}
\affiliation{Hunan Key Laboratory for Micro-Nano Energy Materials
and Devices, Xiangtan University, Hunan 411105, P. R. China;}
\affiliation{School of Physics and Optoelectronics, Xiangtan
University, Xiangtan 411105, China.}
\author{Jin Li}
\affiliation{Hunan Key Laboratory for Micro-Nano Energy Materials
and Devices, Xiangtan University, Hunan 411105, P. R. China;}
\affiliation{School of Physics and Optoelectronics, Xiangtan
University, Xiangtan 411105, China.}
\author{xiangyang Peng}
\affiliation{Hunan Key Laboratory for Micro-Nano Energy Materials
and Devices, Xiangtan University, Hunan 411105, P. R. China;}
\affiliation{School of Physics and Optoelectronics, Xiangtan
University, Xiangtan 411105, China.}
\author{Lijun Meng}
\affiliation{Hunan Key Laboratory for Micro-Nano Energy Materials
and Devices, Xiangtan University, Hunan 411105, P. R. China;}
\affiliation{School of Physics and Optoelectronics, Xiangtan
University, Xiangtan 411105, China.}
\author{Chao Tang}
\affiliation{Hunan Key Laboratory for Micro-Nano Energy Materials
and Devices, Xiangtan University, Hunan 411105, P. R. China;}
\affiliation{School of Physics and Optoelectronics, Xiangtan
University, Xiangtan 411105, China.}
\author{Jianxin Zhong}
\email{jxzhong@xtu.edu.cn}\affiliation{Hunan Key Laboratory for
Micro-Nano Energy Materials and Devices, Xiangtan University, Hunan
411105, P. R. China;} \affiliation{School of Physics and
Optoelectronics, Xiangtan University, Xiangtan 411105, China.}
\date{\today}
\begin{abstract}
In our present work, five previously proposed sp$^3$ carbon crystals
were suggested as silicon allotropes and their stabilities,
electronic and optical properties were investigated by
first-principles method. We find that these allotropes with direct
or quasi-direct band gaps in range of 1.2-1.6 eV are very suitable
for applications in thin-film solar cells. They display strong
adsorption coefficients in the visible range of the sunlight in
comparison with diamond silicon. These five silicon allotropes are
confirmed possessing positive dynamical stability and remarkable
themodynamical stability close to that of diamond silicon.
Especially, the direct band gap M585-silicon possessing
energy higher than diamond silicon only 25 meV per atom
is expected to be experimentally produced for thin-film solar cells. \\
\end{abstract}
\maketitle

\indent Silicon is becoming the fundamental element in our daily
life. Almost all the semiconductor components in high technology of
today are silicon-based. As the second most aboundant element in
earth crust, silicon possesses many advantages not only for
applications in the semiconductor industry. It is also the leading
material foundation in the important field of photovoltaic energy
production. Due to its abundance and ability to translate the solar
energy, as well as the reducing non-renewable traditional energy,
silicon is believed as the future of human. Theoretically, silicon
can forms many allotropes due to its ability of sp$^3$ hybridization
like carbon. But, unfortunately, it is always locked in oxide or lie
in its diamond like ground state.\\
\indent Although the band gap of diamond silicon of about 1.12 eV
\cite{1} lies in the optimal adsorption range of the sunlight, its
indirect characteristic seriously affects its adsorption efficiency
\cite{2}. Thus, many efforts have been paid on searching for viable
silicon allotropes with direct or quasi-direct band gaps overlaping
the sunlight in visible range, for the purpose of improving the
translating efficiency of the solar energy. For example, Botti et
al. have proposed some low-energy silicon allotropes with
quasi-direct band gaps in the range of 0.8-1.5 eV and possessing
strong adsorption coefficients for thin-film solar cell applications
\cite{prb1}. Also, an approach named as Inverse Band Structure
Design Approach (IBSDA) based on the particle swarming optimization
algorithm (PSO \cite{pso}) was developed by Xiang et al. for the
purpose of searching for materials with expected electronic
properties and a quasi-direct band gap silicon allotrope Si$_{20}$-T
was predicted \cite{prl}. Based on the IBSDA as implemented in PSO,
Wang et al. have also found six low energy direct and quasi-direct
silicon allotropes \cite{jacs}. Very recent, Lee et al. suggested
another approach to search for direct band gap silicon allotropes
and they successfully discovered many direct and quasi-direct band
gap allotropes \cite{prb2}. Especially, a recent experimental
progress shows that a cage-like silicon allotrope Si$_{24}$ with a
quasi-direct band gap of about 1.3 eV can be synthesized by
Na-assistanted two step approach \cite{natmat}. The crystal
structure is confirmed as the same as that of the previously
suggested CAS-Si \cite{jpcb}. The authors also show that Si$_{24}$
possesses strong adsorption coefficient higher than diamond silicon
by first-principles calculation \cite{natmat}. All these theoretical
successful examples make us in confident in using
computational-alchemy in the engineering of materials \cite{SA}.\\
\indent In this paper, we predict five low energy silicon allotropes
(M585-silicon, S-silicon, Z-CACB-silicon, H-silicon and
Z-ACA-silicon as shown in Fig. 1) based on the crystal structures of
our previously proposed sp$^3$ carbon phases \cite{hcy1, hcy2} and
find that they are direct or quasi-direct band gap semiconductors
with strong adsorption coefficients for thin-film solar cell
application. Such a transplanting idea of crystal structures have
made a lot of successes in crystal prediction, such as zeolite nets
to carbon allotropes \cite{pccp} and carbon crystals to silicon
allotropes like Bct-C4 silicon \cite{bct,s1}, M4 silicon
\cite{m,s1}, M10 silicon \cite{prb1} and Cco-Si8
\cite{prb1,s2,z1,z2}. Our calculating results show that these five
silicon allotropes possess positive dynamical stability and
remarkable themodynamical stability comparable to that of diamond
silicon. Especially, the direct band gap M585-silicon possessing
energy higher than diamond silicon only 25 meV per atom is expected
to be experimentally produced for thin-film solar cells.\\
\begin{figure}
\center
\includegraphics[width=3.5in]{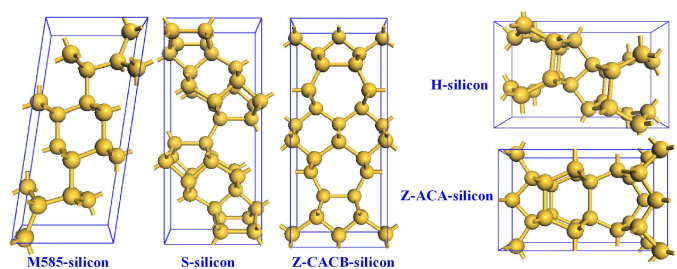}\\
\caption{Perspective view of the optimized crystal structues of
M585-silicon, S-silicon, Z-CACB-silicon, H-silicon and
Z-ACA-silicon.}\label{fig1}
\end{figure}
\begin{table*}
  \centering
  \caption{Total energies, energy band gaps and fundamental structural information including space group, lattice constants and mass densities of diamond silicon, M585-silicon, S-silicon, Z-CACB-silicon, H-silicon and
Z-ACA-silicon.}\label{Tab1}
\begin{tabular}{c c c c c c c c}
\hline \hline
Items             &Diamond silicon         &M585-silicon    &S-silicon       &Z-CACB-silicon  &H-silicon    &Z-ACA-silicon \\
\hline
Space group       &FD-3M (No.227)  &P21/M (No.11)   &CMCM (No.63)    &IMMA (No.74)   &PBAM (No.55)   &PMMN (No.59)\\
Lattice a         &5.407 {\AA}     &14.64 {\AA}     &3.825 {\AA}     &3.833 {\AA}     &11.832 {\AA}   &3.807 {\AA}\\
Lattice b         &5.407 {\AA}     &3.814 {\AA}     &17.141 {\AA}    &7.289 {\AA}     &7.234 {\AA}    &7.055 {\AA}\\
Lattice c         &5.407 {\AA}     &6.772 {\AA}     &7.399 {\AA}     &17.41 {\AA}     &3.812 {\AA}     &12.002 {\AA}\\
Mass density      &2.361 Mg/cm$^3$ &2.311 Mg/cm$^3$ &2.307 Mg/cm$^3$ &2.302 Mg/cm$^3$ &2.284 Mg/cm$^3$ &2.315 Mg/cm$^3$\\
Relative energy   &    0           &25 meV/atom     &42 meV/atom     & 70 meV/atom    &61 meV/atom     &78 meV/atom\\
Band gap          &1.11 eV         & 1.51 eV        & 1.41 eV        & 1.33 eV        & 1.52 eV        & 1.29 eV \\
Direct band gap   &3.32 eV         & 1.51 eV        & 1.53 eV        & 1.38 eV        & 1.63 eV        & 1.43 eV \\
 \hline \hline
\end{tabular}
\end{table*}
\indent In our present work, we perform the structure optimizations
and total energy calculations using the density functional theory
within local density approximation (LDA) \cite{lda1,lda2} as
implemented in Vienna ab initio simulation package (VASP)
\cite{vasp1,vasp2}. The interactions between nucleus and the
3s$^{2}$3p$^{2}$ valence electrons of silicon are described by the
projector augmented wave (PAW) method \cite{paw1,paw2}. A plane-wave
basis with a cutoff energy of 400 eV is used to expand the wave
functions and the Brillouin Zone (BZ) sample meshes are set to be
dense enough (less than 0.21 {\AA}$^{-1}$) to ensure the accuracy of
our calculations. All the crystal structures of silicon allotropes
considered in our present work are fully optimized up to the
residual force on every atom less than 0.005 eV/{\AA}. The dynamical
stabilities of the five new silicon allotropes were evaluated
through simulating their vibrational properties by using the PHONON
package \cite{phonon} with the forces calculated from VASP.
Especially, the hybrid functional method (HSE06) \cite{HSE} is also
considered to accurately calculate the electronic and optical
properties of the five new silicon allotropes.\\
\begin{figure}
\includegraphics[width=3.5in]{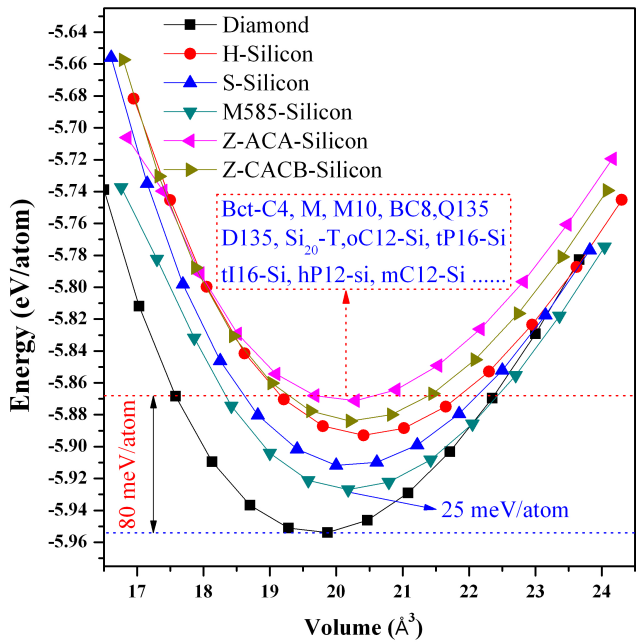}\\
\caption{Energy per atom as a function of volume per atom for
diamond silicon, M585-silicon, S-silicon, Z-CACB-silicon, H-silicon
and Z-ACA-silicon.}\label{fig2}
\end{figure}
\indent As shown in Fig. 1 is the optimized crystal structures of
M585-silicon, S-silicon, Z-CACB-silicon, H-silicon and
Z-ACA-silicon. Their relative energy, energy band gaps, as well as
fundamental structural information including space groups, lattice
constants, mass densities are summarized in Table I , those of
diamond silicon are also listed for comparison. We can see that the
mass density of M585-silicon, S-silicon, Z-CACB-silicon, H-silicon
and Z-ACA-silicon are 2.311 Mg/cm$^3$, 2.307 Mg/cm$^3$, 2.302
Mg/cm$^3$, 2.284 Mg/cm$^3$ and 2.315 Mg/cm$^3$, respectively. These
results indicate that M585-silicon, S-silicon, Z-CACB-silicon,
H-silicon and Z-ACA-silicon are dense close to diamond silicon,
which is good consistent with their sp$^3$ configurations.\\
\indent From the total energies as listed in Table I and shown in
Fig. 2, we find that all these five silicon allotropes are
metastable with energies higher than that of diamond silicon in 80
meV per atom. That is to say, they are themodynamically favorable
than most of the previously proposed silicon allotropes, such as the
quasi-direct band gap Si$_{20}$-T\cite{prl} and Q135\cite{prb2} as
well as the direct band gap D135\cite{prb2} and mC12-Si\cite{jacs}.
In Fig. 2, we also show the total energy per atom as a function of
volume per atom for these new silicon allotropes (Results about
other phase with energies higher than diamond silicon up to 80 meV
per are not shown in it for the considering of simplification). From
the quadratic E-V relation near the equilibrium V$_0$, we can see
that all of them are themodynamically stable protected by obvious
energy barriers in their individual local state. From Table I and
Fig. 2, we can easily know that the most stable one among these five
silicon allotropes is M585-silicon, followed by S-silicon and
H-silicon. Its energy is just 25 meV per atom higher than that of
diamond silicon, which
indicates that it possesses the very high probability to be synthesized in experiment.\\
\begin{figure*}
\includegraphics[width=7in]{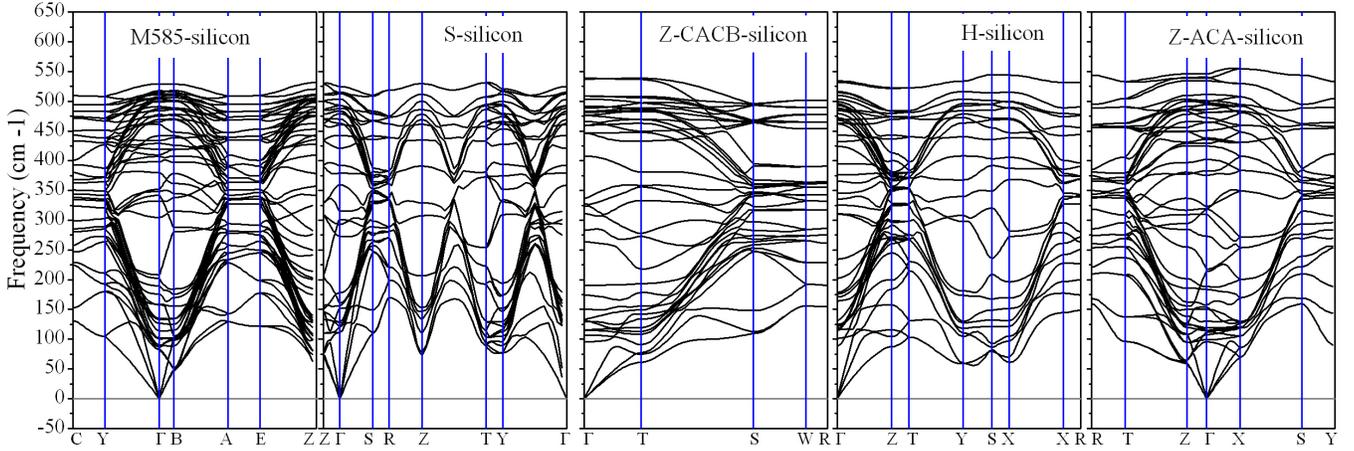}\\
\caption{Phonon band structures of M585-silicon, S-silicon,
Z-CACB-silicon, H-silicon and Z-ACA-silicon.}\label{fig3}
\end{figure*}
\begin{figure*}
\includegraphics[width=7in]{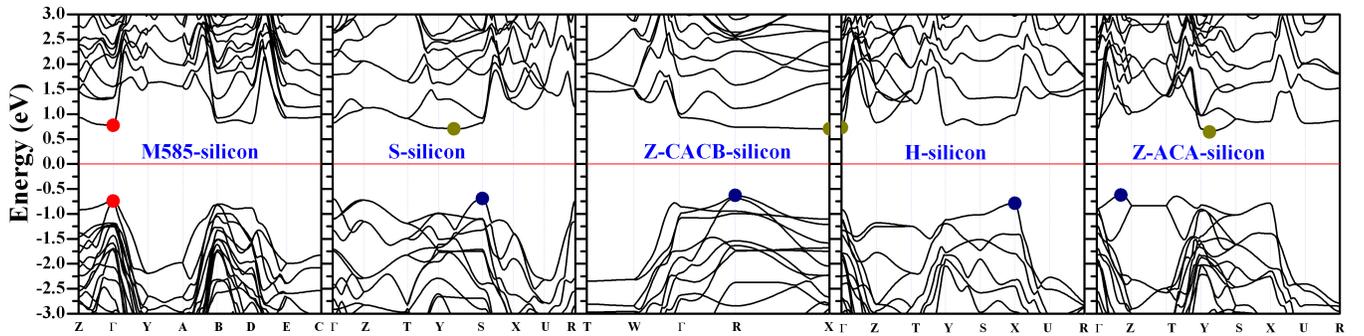}\\
\caption{Electronic band structures of  M585-silicon, S-silicon,
Z-CACB-silicon, H-silicon and Z-ACA-silicon.}\label{fig4}
\end{figure*}
\indent To further confirm the dynamical stabilities of these five
new silicon allotropes, we then investigated their vibrational
properties through simulation their phonon band structures and
density of state. Their phonon band structures are shown in Fig. 3.
According to our results, there is no any imaginary frequency
appearing in these phonon vibrational spectra and no any imaginary
models appearing it their phonon density of state. Such vibrational
properties indicate that these five new silicon allotropes are dynamically stable.\\
\indent As mentioned before, direct or quasi-direct band gap
characteristics are expected for sunlight adsorption, which can
avoiding the requirement of thick absorber layer (for indirect band
gap semiconductors such as diamond silicon) to provide necessary
phonon moment. Thus, we firstly care about the electronic band
structures of these five promising silicon allotropes. Considering
that traditional DFT-method always underestimate the band gap of
semiconductor in comparison of the experiment results, we employ the
HSE06 method in our present work to investigate the electronic
properties of M585-silicon, S-silicon, Z-CACB-silicon, H-silicon and
Z-ACA-silicon. The calculated band structures are shown in Fig. 4
and corresponding band gaps are summarized in Table I. From the band
structures we can see that the most stable M585-silicon is a typical
direct band gap semiconductor with a gap of 1.51 eV. S-silicon,
Z-CACB-silicon, H-silicon and Z-ACA-silicon behave as quasi-direct
band gap semiconductors with band gaps of 1.41 eV, 1.29 eV, 1.33 eV
and 1.52 eV, respectively. Their corresponding direct band gaps are
1.53 eV, 1.38 eV, 1.63 eV and 1.43 eV, respectively. These results
indicate that they are suitable for applications in thin-film solar
cells as absorbers. \\
\begin{figure}
\includegraphics[width=3.5in]{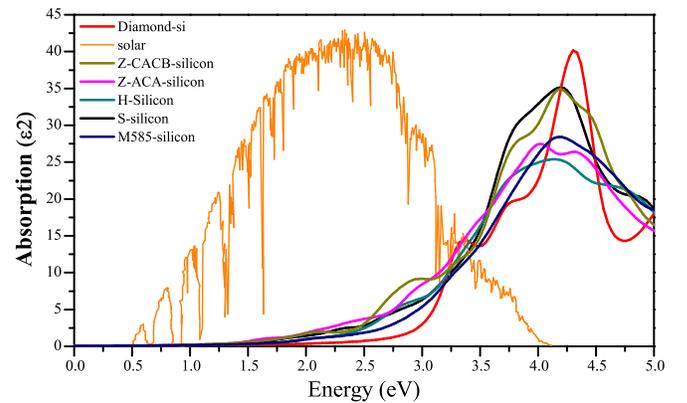}\\
\caption{Adsorption spectra of M585-silicon, S-silicon,
Z-CACB-silicon, H-silicon and Z-ACA-silicon compared to that of
diamond silicon and the reference air mass 1.5 solar spectral
irradiance \cite{solar}.}\label{fig5}
\end{figure}
\indent We then turn our interesting on the absorption abilities of
these five promising silicon allotropes. Absorption spectra of these
allotropes are also calculated based on HSE06 method. The results
are shown in Fig. 5 in comparison with that of the diamond silicon
and the air mass 1.5 solar spectral irradiance \cite{solar}. We can
see that all of these five silicon allotropes possess strong
absorption coefficients better than that of diamond silicon in the
energy range between 1.5 and 3.2 eV. It is well known that indirect
band gap diamond silicon needs necessary phonon moments to assist
its indirect band adsorption \cite{2}. Thus it is always prepared
with thickness about 100 $\mu$m to ensure the adsorption of solar
energy below its direct optical gaps \cite{3}. Our discoveries
suggest that these five new silicon allotropes are suitable for
thin-film solar cell applications in views of their direct or
quasi-direct band gap characteristics and strong adsorption
abilities, if they can be successfully synthesized in future experiment. \\
\indent Usually, silicon is always locked in oxide or lies in its
ground state of diamond silicon. There is no any graphite-like
silicon allotrope that can provides abundant pathways to many other
metastable phases under high pressure condition like carbon
\cite{gtds}. Although diamond silicon can also undergos a series of
phase transitions \cite{trans,t0} with the increase of external
pressure, from diamond to $\beta$-Sn at about 12 Gpa, from
$\beta$-Sn to orthorhombic Imma phase and then to simple hexagonal
at about 13-16 Gpa, from simple hexagonal to an orthorhombic Cmca
phase at about 38 Gpa, from Cmca to a new hexagonal close pack at 42
Gpa and finally to face-centered cubic at 78 Gpa , all these high
pressure phases are metallic and non of them can keep existing at
ambient condition. Fortunately, the process of pressure-release of
$\beta$-Sn phase provides us many opportunities to meet some
metastable silicon allotropes \cite{mets}, such as R8, BC8 and other
unknown phases \cite{t0} with distorted sp$^3$ bondings, depending
on the release velocity and temperature condition. We expect further
experiments in future can find out proper release velocity and
temperature conditions to synthesize these low energy silicon
allotropes based on the high pressure $\beta$-Sn phase. On the other
hand, other experimental methods have also been developed to
synthesize new metastable silicon allotropes \cite{natmat,ex1,ex2}.
M585-silicon, S-silicon, Z-CACB-silicon, H-silicon and Z-ACA-silicon
are topological similar to the recently synthesized allo-Ge phases,
which indicate that they may be synthesized by similar method
\cite{ex2}. The recent success of synthesizing of a cage-like
silicon allotrope Si$_{24}$ with a quasi-direct band gap of about
1.3 eV by Na-assistanted two step approach \cite{natmat} also
provides a suitable method for synthesizing the five new silicon
phases suggested in our present work. Especially, the M585-silicon
with the lowest energy (about 80 meV lower than that of Si$_{24}$)
and cage-like configuration similar to that of Si$_{24}$ is expected
to be synthesized as the
first experimental direct band gap silicon for thin-film solar cell application.\\
\indent In summary, five previously proposed sp$^3$ carbon crystals
were theoretically suggested as silicon allotropes for the first
time and their stabilities, electronic and optical properties were
investigated by first-principles method. We find that these
allotropes with direct or quasi-direct band gaps in range of 1.2-1.6
eV are very suitable for applications in thin-film solar cells. They
display strong adsorption coefficients in the visible range of the
sunlight in comparison with diamond silicon and remarkable
themodynamical stability superior to most of the previously proposed
direct or quasi-direct band gap silicon allotropes such as
Si$_{20}$-T\cite{prl}, Q135\cite{prb2}, D135\cite{prb2} and
mC12-Si\cite{jacs}. Especially, the direct band gap M585-silicon
possessing energy only 25 meV per atom higher than that of diamond
silicon and about 80 meV lower than that of the recently synthesized
Si$_{24}$ \cite{natmat} is expected to be experimentally produced for thin-film solar cells.\\
\indent This work is supported by the National Natural Science
Foundation of China (Grant Nos. A040204 and 11204261), the National
Basic Research Program of China (2012CB921303 and 2015CB921103), the
Hunan Provincial Innovation Foundation for Postgraduate (Grant No.
CX2013A010), the Young Scientists Fund of the National Natural
Science Foundation of China (Grant No. 11204260), and the Program
for Changjiang Scholars and Innovative Research Team in University (IRT13093).\\

\end{document}